\documentclass[letter,amssymb,amsmath,aps,prl,reprint,superscriptaddress,showpacs,floatfix,longbibliography]{revtex4-1}
\usepackage{color}
\usepackage{xcolor}
\usepackage{graphicx}
\usepackage{braket}
\usepackage{tikz}
\usepackage{bm}
\usepackage{hyperref}
\hypersetup{colorlinks=true, citecolor=blue, urlcolor=purple, linkcolor=red}
\usepackage{epstopdf}
\usepackage[space]{grffile}

\newcommand{\liiro}{$\alpha$-Li$_2$IrO$_3$}

\begin{document}
\title{High-resolution resonant inelastic x-ray scattering study of the electron-phonon coupling in honeycomb \liiro}

\author{J. G. Vale}
\email{j.vale@ucl.ac.uk}
\affiliation{London Centre for Nanotechnology and Department of Physics and Astronomy, University College London (UCL), Gower Street, London, WC1E 6BT, United Kingdom}

\author{C. D. Dashwood}
\affiliation{London Centre for Nanotechnology and Department of Physics and Astronomy, University College London (UCL), Gower Street, London, WC1E 6BT, United Kingdom}

\author{E. Paris}
\affiliation{Swiss Light Source, Paul Scherrer Institut, CH-5232 Villigen PSI, Switzerland}

\author{L. S. I. Veiga}
\affiliation{London Centre for Nanotechnology and Department of Physics and Astronomy, University College London (UCL), Gower Street, London, WC1E 6BT, United Kingdom}

\author{M. Garcia-Fernandez}
\affiliation{Diamond Light Source, Rutherford Appleton Laboratory, Didcot, OX11 0DE, United Kingdom}

\author{A. Nag}
\affiliation{Diamond Light Source, Rutherford Appleton Laboratory, Didcot, OX11 0DE, United Kingdom}

\author{A. Walters}
\affiliation{Diamond Light Source, Rutherford Appleton Laboratory, Didcot, OX11 0DE, United Kingdom}

\author{K. J. Zhou}
\affiliation{Diamond Light Source, Rutherford Appleton Laboratory, Didcot, OX11 0DE, United Kingdom}

\author{I.~-M. Pietsch}
\affiliation{Experimentalphysik VI, Center for Electronic Correlations and Magnetism, Augsburg University, 86159 Augsburg, Germany}

\author{A. Jesche}
\affiliation{Experimentalphysik VI, Center for Electronic Correlations and Magnetism, Augsburg University, 86159 Augsburg, Germany}

\author{P. Gegenwart}
\affiliation{Experimentalphysik VI, Center for Electronic Correlations and Magnetism, Augsburg University, 86159 Augsburg, Germany}

\author{R. Coldea}
\affiliation{Clarendon Laboratory, University of Oxford Physics Department,
Parks Road, Oxford, OX1 3PU, United Kingdom}

\author{T. Schmitt}
\affiliation{Swiss Light Source, Paul Scherrer Institut, CH-5232 Villigen PSI, Switzerland}

\author{D. F. McMorrow}
\affiliation{London Centre for Nanotechnology and Department of Physics and Astronomy, University College London (UCL), Gower Street, London, WC1E 6BT, United Kingdom}


\begin{abstract}
The excitations in honeycomb \liiro\ have been investigated with high-resolution resonant inelastic x-ray scattering (RIXS) at the O K edge. 
The low-energy response is dominated by a fully resolved ladder of excitations, which we interpret as being due to multi-phonon processes in the presence of strong electron-phonon coupling (EPC). At higher energies, the orbital excitations are shown to be dressed by phonons. 
The high quality of the data permits a quantitative test of the analytical model for the RIXS cross-section, which has been proposed to describe EPC in transition metal oxides (TMOs).
We find that the magnitude of the EPC is comparable to that found for a range of $3d$ TMOs. This indicates that EPC may be of equal importance in determining the phenomenology displayed by corresponding $5d$ based systems.

\end{abstract}
\maketitle
\section{Introduction}
Understanding the unifying principles describing the myriad of electronic and magnetic phases displayed by transition metal oxides (TMOs) is a central theme of contemporary condensed matter physics.
Recently this challenge has intensified as focus has shifted to the study of $4d$ and $5d$ systems, which are generally less localized than their $3d$ counterparts, while the spin-orbit coupling is significantly enhanced. As the phenomenology of the $4d$ and $5d$ systems has been gradually revealed, questions have naturally arisen concerning the origin  of key similarities and differences. For example, Sr$_2$IrO$_4$ bears many similarities, in terms of structure, magnetism, etc., to the high-temperature superconductor parent compound La$_2$CuO$_4$, and yet has evaded all attempts to induce bulk superconductivity. In this and many other contexts, evaluation of the degree of electron-phonon coupling (EPC) is of special significance.

While estimates of the EPC have long been available from a variety of techniques \cite{pintschovius2005, giustino2008}; it was proposed by Ament \emph{et al.} that resonant inelastic x-ray scattering (RIXS) has in principle several unique advantages \cite{ament_phd, ament2011}; in that it is capable of providing direct information on the EPC that is both element specific and momentum resolved. 
Following this realization, several studies have demonstrated that the EPC for a given phonon mode can indeed be extracted from RIXS data by utilising one of two related methods \cite{lee2013, lee2014, devereaux2016, johnston2016, fatale2016, meyers2018, rossi2019, braicovich2019}. 
Both methods utilise the theoretical RIXS cross-section for phonons as elegantly derived by Ament \emph{et al}.~using a number of standard assumptions (Appendix A). 
The most general approach involves measuring the intensity of the fundamental harmonic (with energy $\hbar\omega_0$) as a function of detuning the incident photon energy away from resonance \cite{rossi2019, braicovich2019}. This method is particularly useful for estimating the EPC in systems where the phonons are poorly resolved due to significant overlap with other excitations, for example, highly damped spin waves; or at absorption edges where the inverse core-hole lifetime $\Gamma$ is larger than the EPC self-energy $M_0$.
At the oxygen K-edge, however, the intermediate state is sufficiently long-lived that multiple phonon overtones $[\hbar\omega_\nu \approx (\nu+1)\hbar\omega_0]$ can be observed \cite{lee2013, lee2014, devereaux2016, johnston2016, fatale2016, meyers2018}. The EPC can be determined at resonance simply through comparing the intensity ratio $I_{\nu}/I_0$ of these overtones to the fundamental.
Here we complement the RIXS studies referenced above by presenting data with significantly better energy resolution than has hitherto been used. This enhanced resolution allows us to obtain data of the highest quality from \liiro, a 5d spin-orbit Mott insulator, thereby facilitating a more stringent test of theory. 

\begin{figure}
\includegraphics{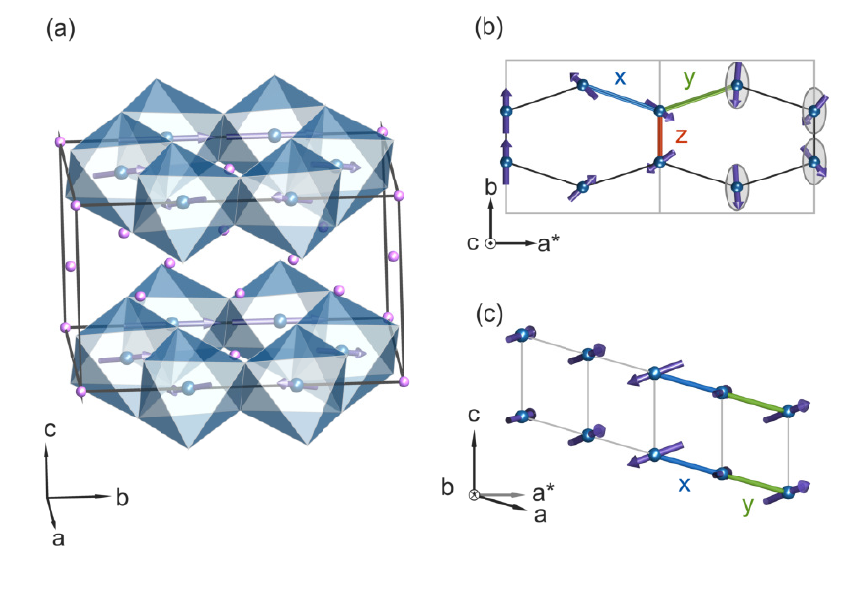}
\caption{(a): Crystal structure of \liiro. The honeycomb 2D networks of IrO$_6$ octahedra are separated by a plane of Li ions. (b): View along the $c$-axis showing magnetic structure and Kitaev bond-directional couplings. Ellipses show the rotation plane of the magnetic moments. (c): View along the $b$-axis showing the ferromagnetic alignment of magnetic moments along the $c$-axis.}
\label{Li213_structure}
\end{figure}

The family of compounds ($\alpha$,$\beta$,$\gamma$)-Li$_2$IrO$_3$, together with Na$_2$IrO$_3$, have been proposed to be proximate to an exotic Kitaev quantum spin liquid state \cite{chaloupka2010, kimchi2011, choi2012, singh2012, katukuri2014, winter2016, winter2017, hermanns2018}. These materials contain a network of edge-sharing IrO$_6$ octahedra, which in the case of \liiro, form a honeycomb arrangement in the $ab$ plane \cite{omalley2008} [Fig.~\ref{Li213_structure}(a)].
The ideal Kitaev spin liquid state requires a perfect $90^{\circ}$ Ir-O-Ir bond angle \cite{jackeli2009, winter2017}; however weak lattice distortions for all polymorphs of ($\alpha$,$\beta$,$\gamma$)-Li$_2$IrO$_3$ and Na$_2$IrO$_3$ discovered thus far give rise to additional non-Kitaev exchange terms in the Hamiltonian, and hence, long-ranged magnetic order \cite{choi2012, biffin2014_beta, takayama2014, biffin2014_gamma, williams2016, choi2019} [Fig.~\ref{Li213_structure}(b,c)].
An outstanding question is the role that lattice coupling plays in such Kitaev-type materials. For instance, Raman scattering data on ($\beta$,$\gamma$)-Li$_2$IrO$_3$ reveals Fano-type behavior characteristic of strong coupling to a continuum of spin excitations \cite{glamazda2016}. Various recent works on other almost ideal Kitaev systems have also proposed unconventional thermal transport behavior \cite{nasu2017}, thermal Hall effect phenomenology \cite{vinkleraviv2018}, and significant spin-phonon scattering \cite{hentrich2018}.

Here we show that by using high-resolution RIXS at the O K-edge it is possible to extract unique information about the EPC in \liiro. This includes its momentum dependence, the existence of possible anharmonicity, and phonon dressing of the orbital excitations.
In general our results establish the fact that the EPC can be as significant for 5$d$ systems as for 3$d$ ones, while more specifically highlighting the importance of coupling between the electronic and lattice degrees of freedom in Kitaev systems.

\begin{figure}
\includegraphics{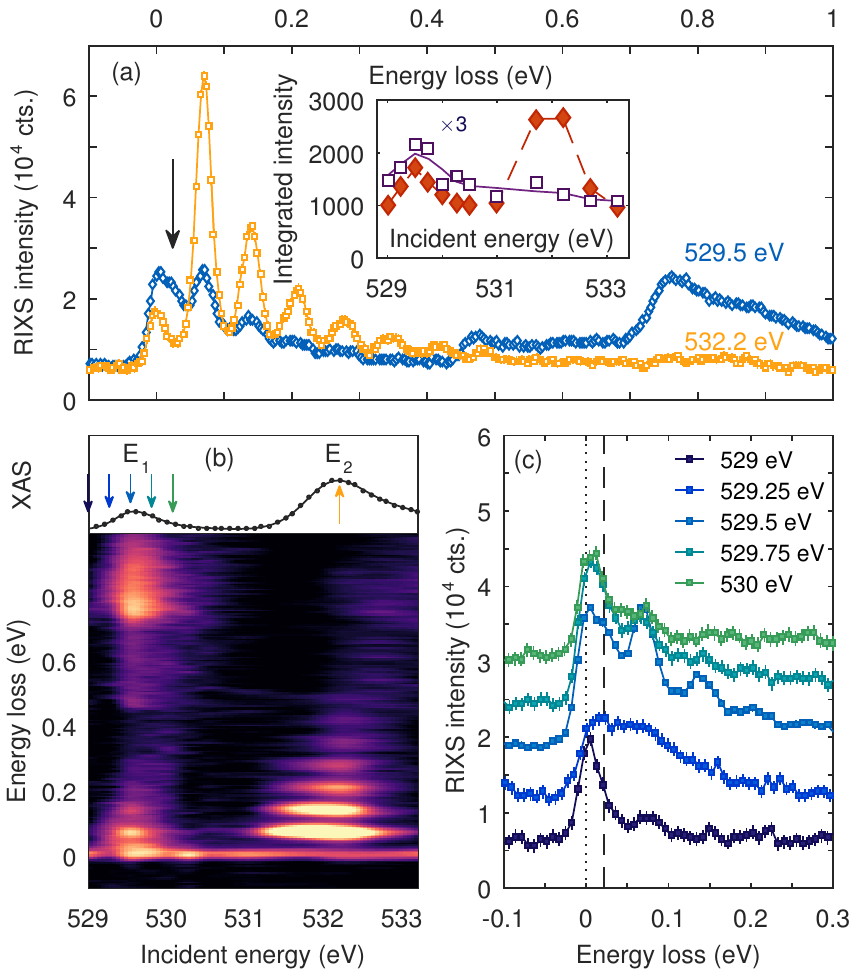}
\caption{Incident energy dependence of the experimental RIXS cross-section at 10~K for $\theta = \text{25}^{\circ}$ (equivalent to an in-plane momentum transfer $Q_{||}=\text{0.43 \AA}^{-\text{1}}$, for details see Table \ref{hkl_table}). (a): Representative spectra for incident energies of 529.5~eV (blue) and 532.2~eV (yellow). The arrow highlights the feature at 20~meV energy loss. The inset shows the integrated intensity for $\text{10}\leq E (\text{meV}) \leq \text{35}$ (open squares) and $\text{40} \leq E (\text{meV}) \leq \text{100}$ (filled diamonds) as a function of incident energy, highlighting the different resonant behavior for these features. (b): RIXS map and x-ray absorption spectrum (linear horizontal incident polarization). Arrows show the incident energies used for the cuts plotted in the rest of the figure. (c): RIXS spectra plotted for various incident energies. Dotted line: zero energy loss. Dashed line: Low energy feature at 20~meV energy loss.}
\label{RIXS_map}
\end{figure}
\section{Methods}
High-resolution oxygen K-edge (530~eV) RIXS measurements were performed on the I21-RIXS spectrometer at Diamond Light Source. A single crystal of \liiro\ (300 $\times$ 150 $\times$ 100 $\mu\text{m}^3$, crystal structure plotted in Fig.~\ref{Li213_structure}) was mounted with silver epoxy onto a copper sample holder, and oriented such that the a* and c* reciprocal lattice axes lay in the horizontal scattering plane [Fig.~\ref{XAS}(a,b)].
Note twinning of samples of $\alpha$-Li$_2$IrO$_3$ is common due to the monoclinic crystal symmetry and similarity in magnitude of the $a$- and $c$-axis lattice parameters.
The sample was cleaved in ultra-high vacuum ($<\!10^{-8}$ mbar) and placed onto a six-axis sample manipulator \emph{in situ}. Throughput was increased by the presence of a collecting mirror close to the sample position. The scattering angle was fixed at $2\vartheta=\text{154}^{\circ}$, with the scattered photons discriminated by an Andor CCD detector. Unless otherwise stated, all measurements were performed with linear horizontal ($\pi$) incident polarization.
The total energy resolution was determined to be 26~meV ($E/\Delta E \sim 20000$) based on quasielastic scattering from carbon tape placed close to the sample; roughly a factor of two better than previous studies performed at the O K edge \cite{meyers2018, lu2018}. 
All RIXS spectra were corrected for self-absorption using the method described in Appendix \ref{selfabs}.
Experimentally, it was found that the sample was miscut with respect to the $\theta$-rotation axis; with the specular condition satisfied for $\theta \sim 80^{\circ}$. 
Assuming that the $b^*$-axis points vertically out of the scattering plane, then the corresponding Miller indices for different values of $\theta$ and $2\vartheta=\text{154}^{\circ}$ are given in Table~\ref{hkl_table}.

\begin{figure*}
\includegraphics{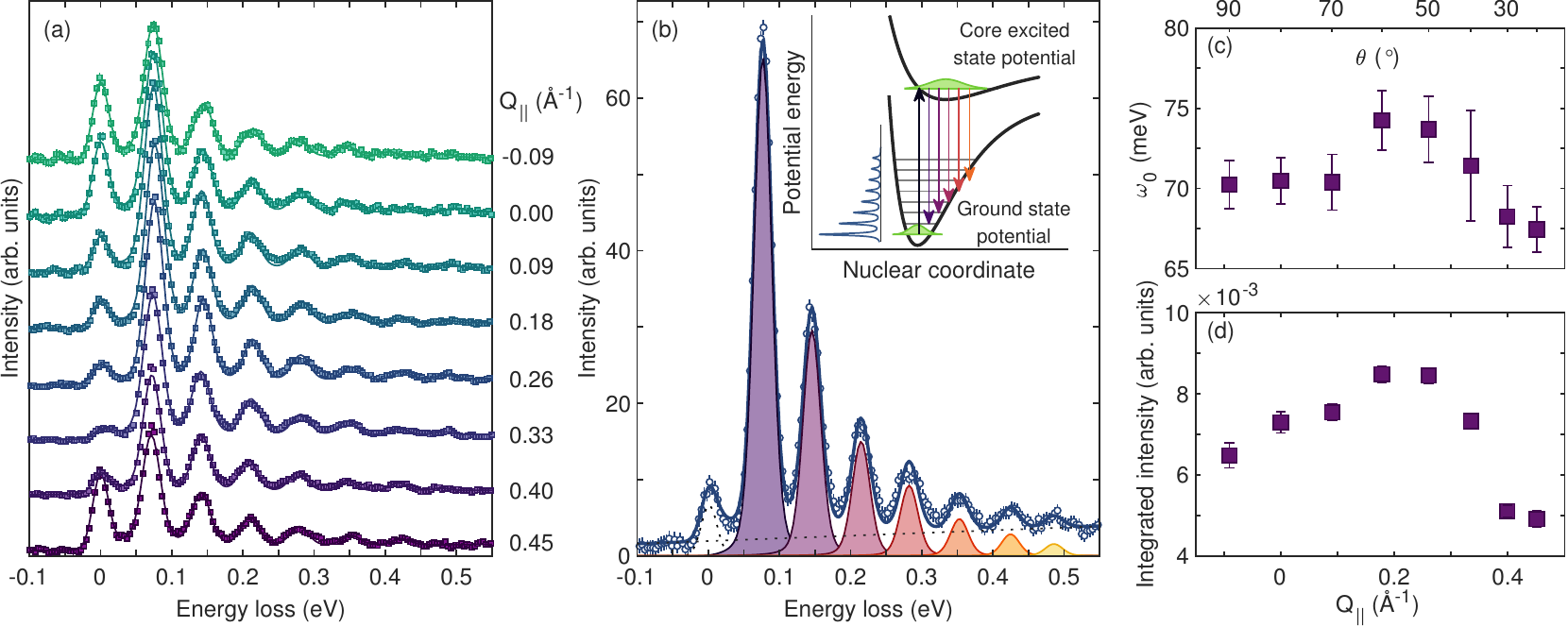}
\caption{(a): Momentum dependence of the low-energy portion of the RIXS spectra collected for $E_{in}=E_2$. (b): Representative fit for $Q_{||} = \text{0.26\,\AA}^{-1}$
($\theta = \text{50}^{\circ}$). Shaded peaks highlight the multiphonon excitations, while the dashed line indicates the excitation at 20~meV. Inset details the process by which multiphonon excitations can be measured. (c,d): Energy (c) and integrated intensity (d) of fundamental phonon as obtained from least squares fitting.}
\label{Qdep_10K}
\end{figure*}
\begin{figure}
\includegraphics[scale=1]{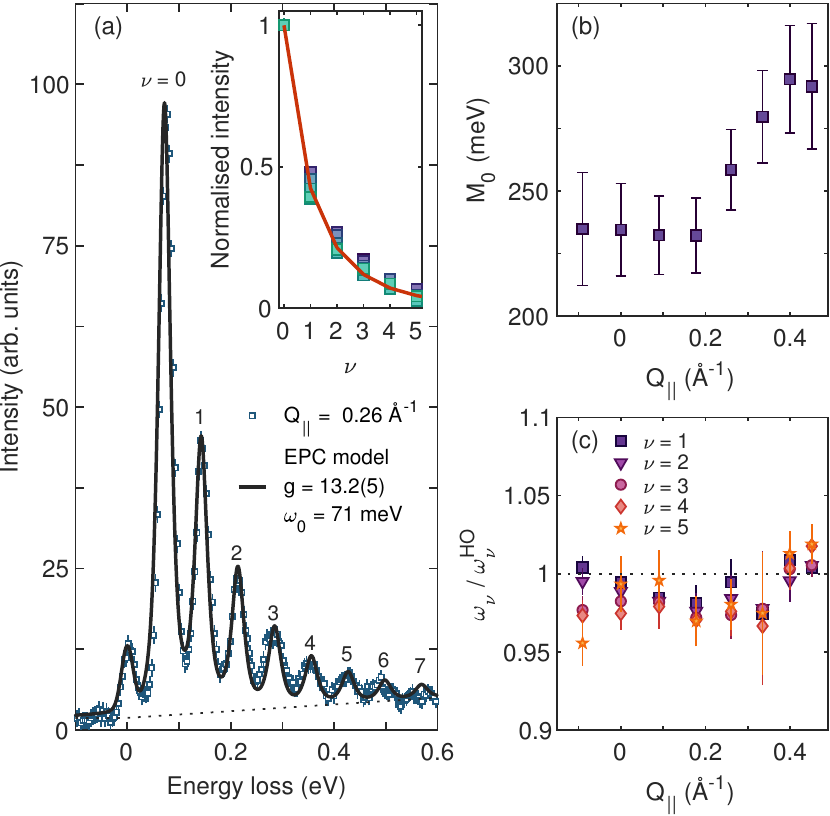}
\caption{(a): Representative RIXS spectrum collected with $E_i = E_2$. Solid line is the best fit to the theoretical RIXS cross-section as described in the main text. Inset shows intensities of phonon harmonics as a function of $Q_{||}$; 
solid line is the best global fit to the Ament model with $\omega_0 = \text{71(3)~meV}$, $g=(M_0/\omega_0)^2=14(3)$. (b): Extracted electron-phonon coupling $M_0$ as a function of $Q_{||}$. (c): Relative deviation of the frequency of individual phonon modes $\omega_n$ from the value expected for an ideal harmonic oscillator $\omega^{HO}_n$.}
\label{EPC}
\end{figure}

\section{Results}
In Fig.~\ref{RIXS_map}, we present O K-edge RIXS spectra collected at 10~K as a function of incident photon energy $E_{in}$ for fixed incidence angle $(\theta=\text{25}^{\circ})$.  As expected, the resonant scattering cross-section is dramatically increased at the maxima of the pre-edge features $E_1$ (529.5~eV) and $E_2$ (532.2~eV) in the x-ray absorption spectrum [Figs.~\ref{RIXS_map}(b,c)]. The peaks at $E_1$ and $E_2$ arise due to hybridization between the Ir $5d$ $t_{2g}$ ($e_g$) and O $2p$ states respectively.  
There are, however, significant differences between RIXS spectra collected at those incident energies; as highlighted in Fig.~\ref{RIXS_map}(a). 
The broad excitations which are evident above 0.4~eV energy loss for $E_{in} = E_1$ are strongly suppressed at $E_2$. Comparison with previously obtained data at the Ir L$_{\text{3}}$ edge shows that these features are consistent with $j_{3/2}\leftarrow j_{1/2}$ excitations at the Ir site \cite{gretarsson2013, bhkim2014, revelli2019}. This includes a sharp feature at 0.46~eV which originates from coupling between the electron-hole excitation of the $j_{\text{eff}}=1/2$ band, and the ``spin-orbit exciton'' close to the optical absorption edge \cite{bhkim2014}.
The fact that they can be seen at the O K-edge is due to the aforementioned Ir-O hybridization, with similar behavior observed previously for the perovskite iridates \cite{lu2018}.

The most prominent feature in Fig.~\ref{RIXS_map}(a), however, is a series of peaks which have approximately equal separation ($\omega_0 \sim$ 70~meV) and monotonically decrease in intensity with increasing energy loss. These peaks are more pronounced for $E_{in} = E_2$, but are clearly present at both incident energies. 
We propose that these peaks are an approximately harmonic progression of multiphonon processes ($\nu=0,1,2,\dots,N$). It has been previously proposed in the literature that these manifest as a result of significant EPC in the ground state \cite{ament2011, devereaux2016}.
At the O K-edge, one would heuristically expect to be most sensitive to phonon modes involving a significant displacement of oxygen atoms. Our value for $\omega_0$ is in excellent agreement with a purely oxygen-based $A_u$ phonon mode at 72~meV, which has previously been observed by optical conductivity \cite{hermann2017}. We therefore assign it accordingly.

One of the key advantages of RIXS compared to other spectroscopic techniques is the ability to perform measurements as a function of momentum transfer. 
In Fig.~\ref{Qdep_10K}(a), we present data collected at $E_{in}=E_2$ for different values of $Q_{||}$.
We first focus upon the excitations below 0.4~eV, which clearly show the multi-phonon processes discussed previously. The data were fitted with multiple pseudo-Voigt functions on a linear sloping background. The Gaussian component widths were constrained to the resolution width ($\Delta E=\text{26~meV}$). The $\nu\!=\!0$ phonon at $\omega_0$ is found to be weakly dispersive, with a concurrent slight variation of the RIXS intensity [Fig.~\ref{Qdep_10K}(c,d)], as expected for an optical phonon at low $\mathbf{q}$ \cite{devereaux2016}. 
As discussed previously, one can determine the EPC strength as a function of momentum transfer,
simply by modelling the experimental data with the theoretical RIXS cross-section provided in Ref.~\onlinecite{ament2011}. This approach is henceforth referred to as the Ament model.
By constraining the magnitude of the inverse corehole lifetime $\Gamma=\text{0.235~eV}$ (extracted from fits to X-ray absorption spectra, see Appendix \ref{XAS_appendix}), and $\omega_0$ to the experimental value obtained by RIXS, then the only remaining free parameters are the EPC strength $M_0$, and an overall scale factor.
We show the results of a fit to this model in Fig.~\ref{EPC} for $E_{in} = E_2$, that is, at resonance.
The Ament model provides an excellent description of the intensities of the phonon satellites out to $\nu=5$ [Fig.~\ref{EPC}(a)].
We find that the magnitude of the EPC is weakly dependent upon momentum transfer [Fig.~\ref{EPC}(b)], with a mean value of
 $M_0 = \text{260(30)~meV}$, equivalent to $g=M_0^2/\omega_0^2=14(3)$ in dimensionless units.
A similar magnitude of EPC is obtained from fits to data collected with $E_{in}=E_1$ (see Appendix). 
To our knowledge, this is the first estimate of the EPC strength in a bulk 5d transition metal oxide.
In principle, the magnitude and momentum dependence of the EPC should vary for different phonon modes \cite{braicovich2019}. We stress that the value obtained experimentally is for this purely oxygen-based $A_u$ mode.

\begin{figure}
\includegraphics{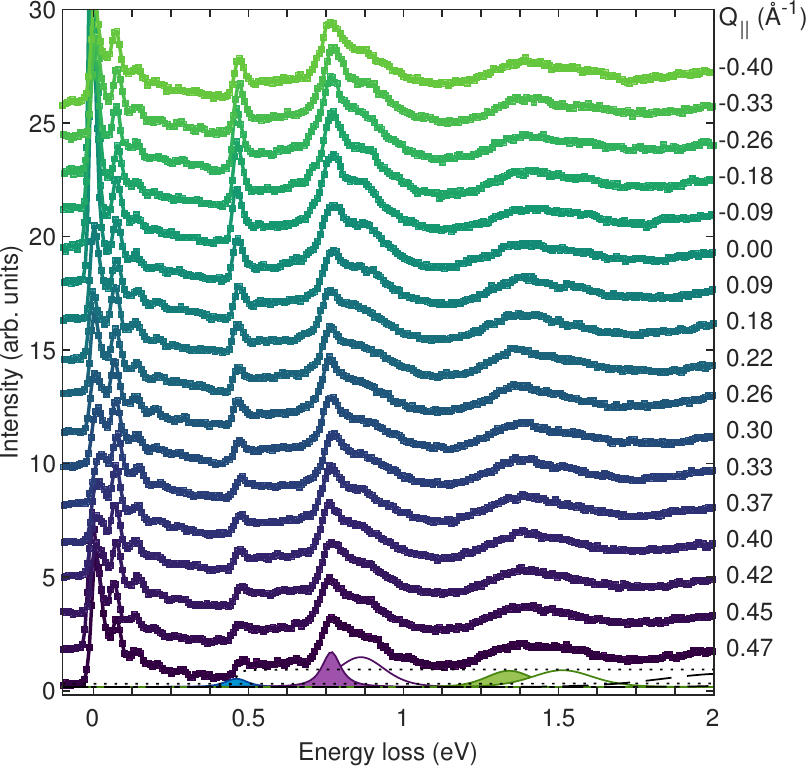}
\caption{Momentum dependence of RIXS spectra collected at 10~K ($E_{in}=E_1$). Highlighted peaks are the spin-orbit exciton (blue), local $j_{3/2}\leftarrow j_{1/2}$ transitions (purple), and intersite transitions (green).}
\label{EPC_dd}
\end{figure}

\begin{figure*}
\includegraphics{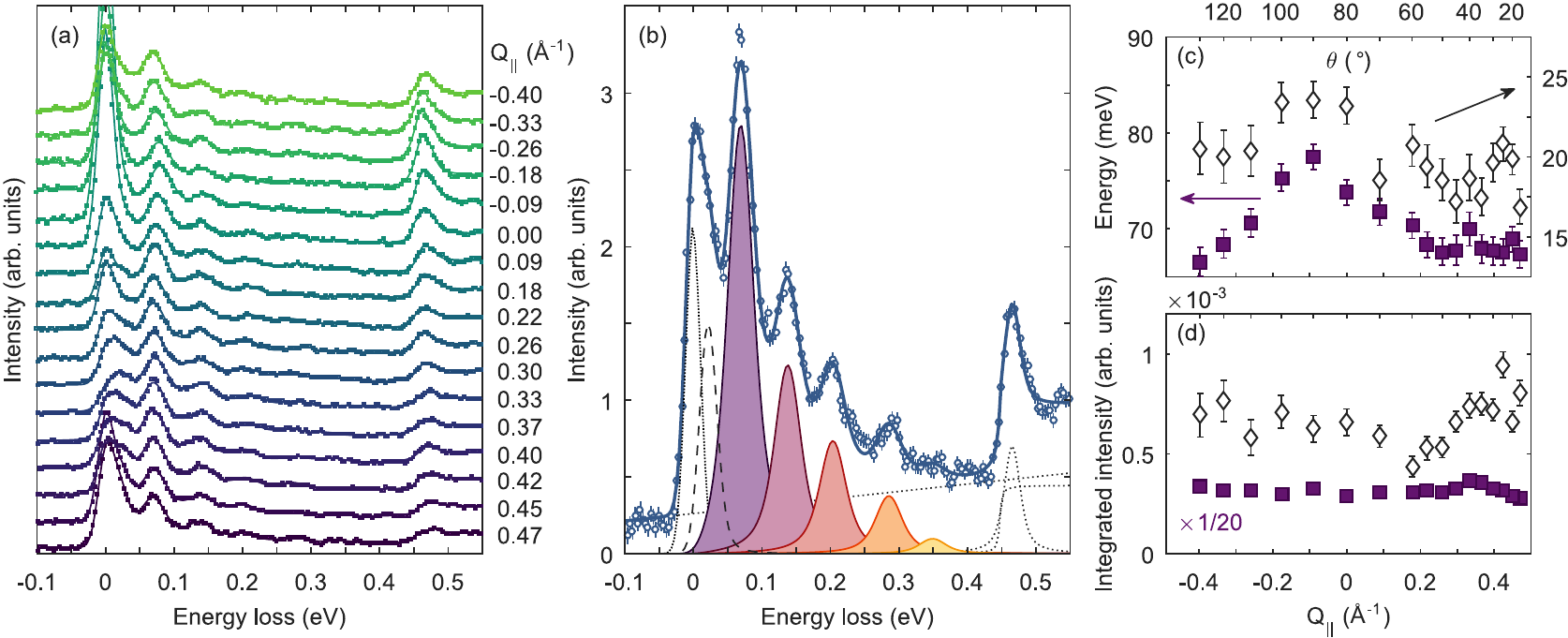}
\caption{(a): Momentum dependence of the low-energy portion of the RIXS spectra collected for $E_{in}=E_1$. (b): Representative fit for $Q_{||} = \text{0.26\,\AA}^{-1}$
($\theta = \text{50}^{\circ}$). Shaded peaks highlight the multi-phonon excitations, while the dashed line indicates the excitation at 20~meV. (c,d): Energy (c) and integrated intensity (d) of fundamental phonon (filled squares) and 20~meV (empty diamonds) excitations as obtained from least squares fitting.}
\label{Qdep_10K_E1}
\end{figure*}

At $E_{in}=E_2$, one probes the effect of hybridization between the oxygen $2p$ orbitals and the unoccupied Ir $5d$ $e_g$ states. However, a lot of the interesting physics -- including the $j_{\text{eff}}=1/2$ ground state -- involves the \emph{occupied} 5d $t_{2g}$ orbitals. In Figures \ref{EPC_dd} and \ref{Qdep_10K_E1}(a), we present RIXS spectra collected at 10~K as a function of momentum transfer, with incident energy $E_{in}=E_1$. The two figures focus on the orbital and low-energy excitations respectively.

\begin{figure}
\includegraphics{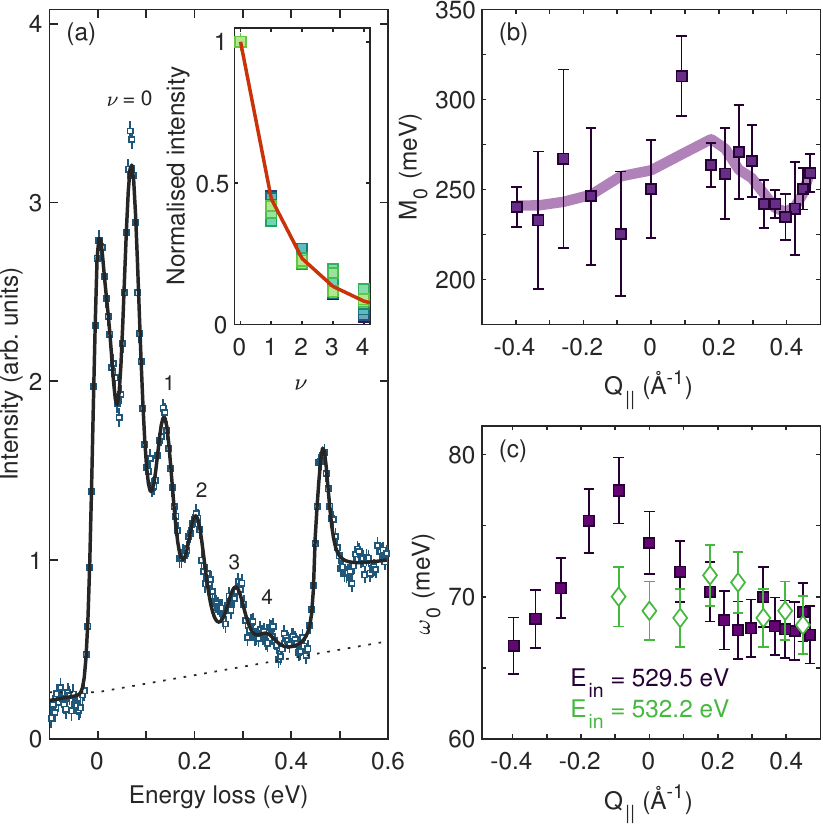}
\caption{(a): Representative RIXS spectrum collected with $E_i = E_1$. Solid line is the best fit to the theoretical RIXS cross-section as described in the main text. Inset shows intensities of phonon harmonics as a function of $Q_{||}$; 
solid line is the best global fit to the Ament model with $\omega_0 = \text{71(3)~meV}$, $g=(M_0/\omega_0)^2=14(3)$. (b): Extracted electron-phonon coupling $M_0$ as a function of $Q_{||}$. Solid line is a guide to the eye. (c): Comparison of the phonon dispersion at $E_i=E_1$ (filled squares) and $E_{in}=E_2$ (open diamonds).}
\label{EPC_529eV}
\end{figure}

We start with the low-energy excitations below 0.6~eV. As in the previous section, the data were fitted with multiple pseudo-Voigt functions on a linear sloping background, along with an arctangent step function representing the electron-hole continuum above 0.4~eV. 
Fitting of the low-energy excitations reveals that the phonon is weakly dispersive [Fig.~\ref{Qdep_10K_E1}(b--d)], with a similar energy to that observed for $E_{in}=E_2$.
The multi-phonon harmonics can be well-described by the Ament model (Fig.~\ref{EPC_529eV}), with the mean EPC self-energy $M_0=\text{250(20)~meV}$ also comparable in magnitude to the value at $E_2$. This is somewhat surprising given that the coupling between orbitals is different at the two incident energies.
Nevertheless, there are some differences. The form of the phonon dispersion appears qualitatively different, especially for $Q_{||}\rightarrow 0$ [Fig.~\ref{EPC_529eV}(c)]. Moreover, the multi-phonon peaks are noticeably broader at $E_{in}=E_1$ compared to $E_2$. These observations suggest the presence of two phonon modes which have different resonant behavior due to separate origins. 
It is not possible, unfortunately, to conclusively determine within the limits of our data.
Finally it was found that the quality of the fits were noticeably improved with the inclusion of an additional low-energy peak at 20~meV; we discuss its origins later.

\begin{figure}
\includegraphics{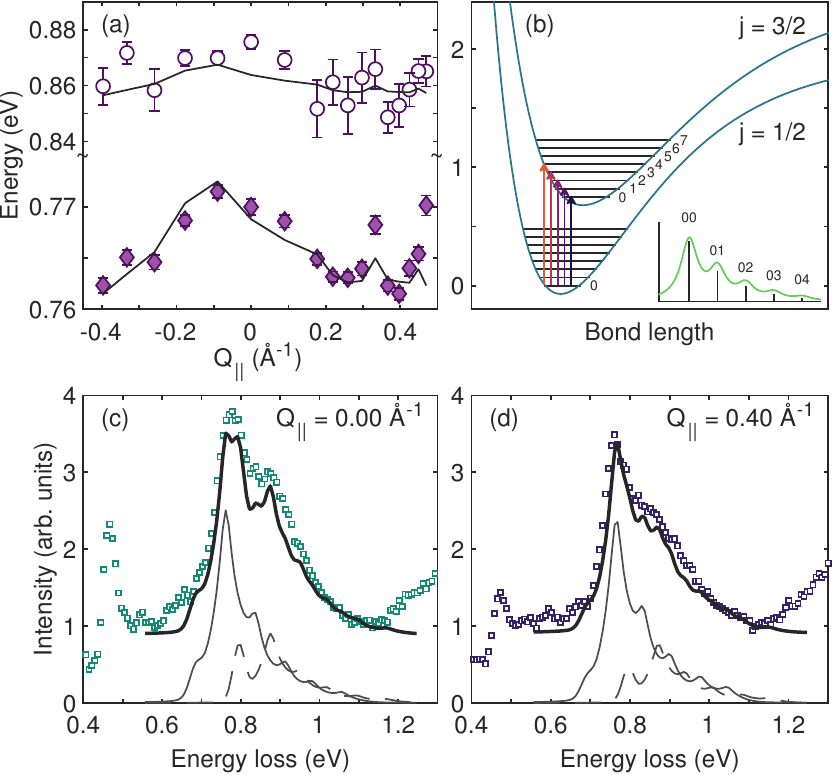}
\caption{(a): Fitted energies of the $j_{3/2}\leftarrow j_{1/2}$ excitations. The overlaid solid lines are the momentum dependence of $\omega_0$ which have been shifted upwards in energy [Fig.~\ref{Qdep_10K}(c)]. (b): Overview of the Franck-Condon process. For clarity only one of the $j_{3/2}$ bands has been shown. Numbers next to the $j_{3/2}$ band refer to the phonon harmonic $\nu$. Inset shows the intensity of vibronic satellites as a function of energy. (c,d): Orbital excitations for $Q_{||}=\text{0.0\,\AA}^{-1}$ (c) and $Q_{||}=\text{0.4\,\AA}^{-1}$ (d) at 10~K. Overlaid is the best fit to the peak profile, which comprises two $dd$-excitations that have both been dressed by phonons as a consequence of EPC. The relevant parameters $\omega_0$, $g$ are constrained to the values obtained for fits at lower energies (Fig.~\ref{EPC}).}
\label{EPC_dd_fits}
\end{figure}

Now that the behavior of the low-energy excitations has been well established, we turn to the higher energy orbital transitions (Figs.~\ref{EPC_dd}, \ref{EPC_dd_fits}). In Fig.~\ref{EPC_dd_fits}(a), we present the results of fits to the $j_{3/2}\leftarrow j_{1/2}$ excitations at $\approx\text{0.8~eV}$ energy loss. They appear to exhibit a similar dispersion to the single phonon excitation at $\omega_0 = \text{70~meV}$.
This can be explained if the orbital excitations are considered to be dressed by phonons.
Consider an electronic transition from the $j_{1/2}$ ground state to the $j_{3/2}$ state. Since $\hbar\omega_0 << k_B{T}$, we assume that all transitions originate from the $\nu=0$ vibrational level in the ground state. 
The transition probability is directly proportional to the overlap between the vibrational wavefunctions in the ground and excited states (Franck-Condon principle). 
In principle, this means that it should be possible to observe so-called vibronic satellites $j_{3/2}^{\nu=N}\leftarrow j_{1/2}^{\nu=0}$ off the fundamental orbital excitation $j_{3/2}^{\nu=0}\leftarrow j_{1/2}^{\nu=0}$ [Fig.~\ref{EPC_dd_fits}(b)] \cite{rothamel1983, liu2003, hancock2010}. Similar excitations have previously been observed with oxygen K-edge RIXS for diatomic oxygen \cite{hennies2010}, and the one-dimensional cuprate Li$_2$CuO$_2$ \cite{johnston2016}.

We constructed a simple model in which the purely electronic $j_{3/2}^{\nu=0}\leftarrow j_{1/2}^{\nu=0}$ excitation is dressed by vibronic satellites, which are described by the theoretical RIXS cross-section detailed by Ament et al \cite{ament2011}. The values for the fundamental phonon energy $\omega_0$ and dimensionless electron-phonon coupling parameter $g=(M_0/\omega_0)^2$ were constrained at each momentum transfer to the values obtained from the fits at lower energy loss.
We find that our model provides a good description of the experimental data [Figs.~\ref{EPC_dd_fits}(c,d)]. One notable feature is that the $j_{3/2}^{\nu=0}\leftarrow j_{1/2}^{\nu=0}$ transitions are of lower intensity than the equivalent $j_{3/2}^{\nu=1}\leftarrow j_{1/2}^{\nu=0}$ process. This is characteristic behavior for an electronically forbidden ($\Delta S \neq 0$), but vibronically allowed, transition, in which there is a large change in geometry between the ground and excited states.

\begin{figure}
\includegraphics{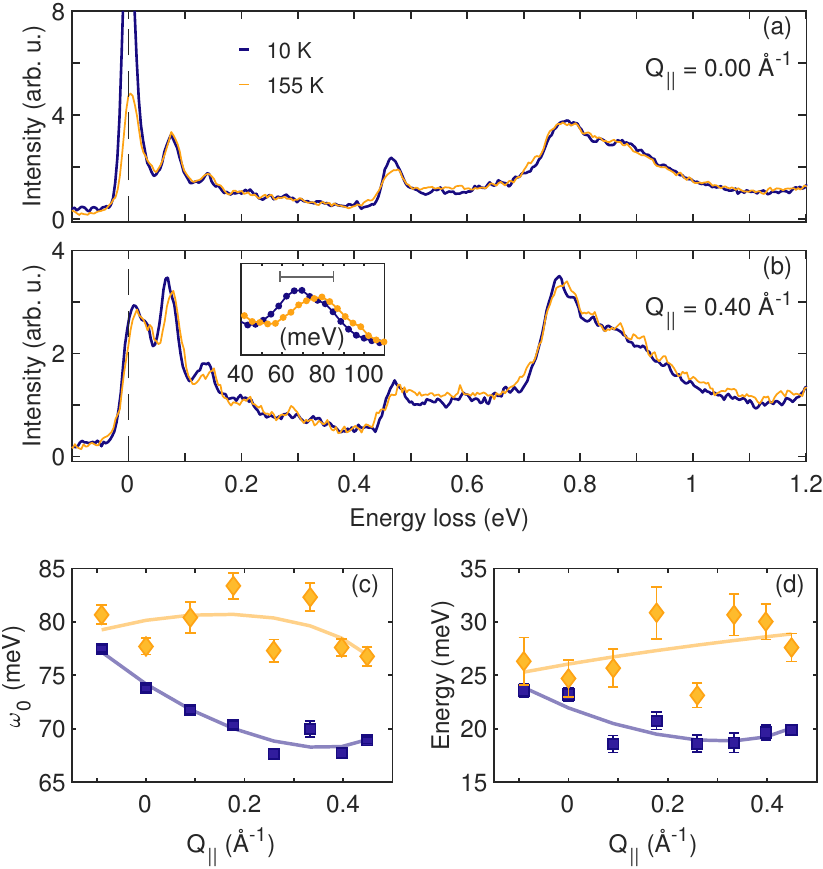}
\caption{(a,b): Comparison of RIXS spectra collected at 10~K (purple) and 155~K (yellow) for an in-plane momentum transfer $Q_{||} = \text{0, 0.4 \AA}^{-1}$. Inset in (b) shows shift of first phonon peak with temperature; scale bar is the instrumental resolution. (c,d): Dependence of fundamental phonon frequency $\omega_0$ (c) and low-energy feature (d) upon $Q_{||}$. Solid lines are guides to the eye.}
\label{Tdep}
\end{figure}

The models we have used to describe the electron-phonon coupling assume ideal, non-dispersive, harmonic oscillators. However, the lattice interaction potential in real materials is anharmonic. Anharmonicity can occur as a consequence of EPC and/or phonon-phonon interactions, and has two important ramifications for \liiro.
The first is that higher-order phonon harmonics are progressively lower in energy than would be expected for an ideal harmonic oscillator ($\hbar\omega_{\nu} < \hbar\omega_{\nu}^{\text{HO}}$). This broadly appears to be the case at 10~K within experimental uncertainty [Fig.~\ref{EPC}(d)].

Anharmonic effects also generally result in phonon softening at high temperatures. To this end, we collected RIXS data at 155~K -- well above the N\'{e}el temperature $T_N=\text{15~K}$, and $E_{in}=E_1$. Representative spectra are presented in Figs.~\ref{Tdep}(a) and (b) for two different values of $Q_{||}$. We note two relevant differences as a function of temperature. 
The first is that some broadening can be observed above $T_N$ around the spin-orbit exciton at 0.4~eV. A number of studies on other iridates have suggested that the existence and behavior of this mode is intrinsically linked to the magnetic order present in the system \cite{kim2012_sr214, gretarsson2013}. Therefore the broadening of the spin-orbit exciton may be indirect evidence for the loss of long-ranged magnetic order in \liiro.

It also appears that $\omega_0$ significantly hardens below $T_N$, with the maximum hardening occurring close to the magnetic ordering wavevector \cite{williams2016}. This is contrary to the expectations for a conventional anharmonic oscillator, and may be evidence for spin-phonon coupling at low temperatures, putatively to low-energy spin excitations \cite{knolle2014, glamazda2016, choi2019}. 
One further possibility that we are unable to rule out is that there are in fact two overlapping phonon progressions, each with slightly different $\omega_0$. Whether this is due to coupling to two different phonon modes, or the result of sample twinning, is not possible to determine conclusively within the limits of the instrumental resolution. Multiple phonon progressions would also explain the increased broadening of the higher-order satellites; however, this can also be explained by a number of sources, including dispersion of the optical phonon, or EPC itself.

\begin{figure}
\includegraphics{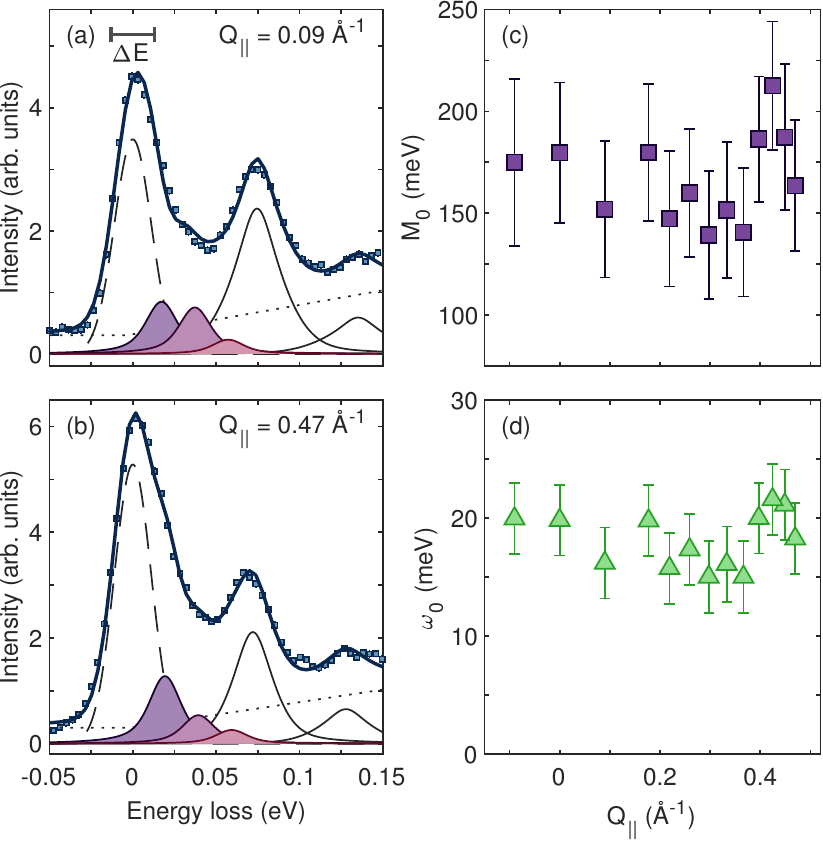}
\caption{(a,b): Low energy part of RIXS spectra collected at $E_{in}=E_1$, 10~K, and $Q_{||} = \text{0.09}$, $\text{0.47~\AA}^{-1}$. Overlaid is the best fit to the data, along with the respective components. Filled peaks correspond to harmonics of phonon at $\approx\text{20~meV}$. (c). Fitted value of EPC self-energy $M_0$. (d): Fitted value of fundamental phonon energy $\omega_0$.}
\label{lowE_harmonics}
\end{figure}

In the low-temperature data, we also observed a weak, dispersive, peak at $\sim \text{20~meV}$, which appears to resonate at $E_{in} = E_1$ [inset of Fig.~\ref{RIXS_map}(a)]. This is indicative of a process involving Ir valence states.
A similar feature has recently been observed at the Ir $L_3$ edge by Revelli \emph{et al} \cite{revelli2019}; who propose that it manifests from scattering from a broad continuum of spin excitations indicative of proximity to a Kitaev spin liquid state.
Theoretical calculations of the RIXS cross-section predict, however, that this spin continuum should have a small RIXS cross-section in the vicinity of the $\Gamma$ point \cite{halasz2016}. Kinematic constraints should therefore limit its observability at the oxygen $K$-edge.
We note that the energy scale of the peak is consistent with what would be expected for an Ir-O stretching mode \cite{glamazda2016}. If this is the case, then in theory one should also observe multi-phonon harmonics for this mode. Such behavior has been previously demonstrated to be relevant for Na$_2$IrO$_3$ \cite{gretarsson2013_prb}.
The RIXS data below 0.15~eV were fitted with multiple pseudo-Voigt functions representing the elastic line, three harmonics of the 20~meV phonon, and two harmonics of the 70~meV phonon. The Gaussian component of these peaks was fixed to the instrumental resolution width, and a sloping background was added to account for higher energy excitations.
We present the results in Fig.~\ref{lowE_harmonics}, which shows that the data are well described by this model [Fig.~\ref{lowE_harmonics}(a,b)].
By fitting the amplitudes of the harmonics using the Ament model, we obtain a very large mean value for the dimensionless coupling $g$ [$g=\text{87(6)}$], which corresponds to a self-energy $M_0=\text{170(20)~meV}$ [Fig.~\ref{lowE_harmonics}(c,d)]. 

\section{Discussion}
Our experimental data clearly establish that electron-lattice coupling is important for \liiro. In order to put this into context, we compared our experimentally obtained value for $M_0$ to those obtained for other TMOs via RIXS (Table \ref{EPC_summary}).
$M_0$ appears to be of broadly similar magnitude across a wide range of systems, regardless of oxidation state, coordination geometry, or size of $\omega_0$. 
This implies that EPC should be treated on an equal footing both for 3$d$ and 5$d$ TMOs.
One word of caution is that the absolute magnitude of $M_0$ extracted from any fit to the EPC model used above is strongly dependent on the choice of the inverse corehole lifetime $\Gamma$. 
For instance, the calculated amplitudes are practically identical if $\Gamma_1=\text{0.24~eV}$, $g_2 = 14(3)$; or $\Gamma_2=\text{0.15~eV}$, $g_2=g_1(\Gamma_2/\Gamma_1)^2=5.7(13)$. Here $\Gamma_1$ and $\Gamma_2$ refer respectively to the value obtained from our XAS data (Appendix \ref{XAS_appendix}), and theoretical results \cite{coreno1999}.
One recent suggestion by Geondzhian and Gilmore \cite{geondzhian2018} is that rather than electron-phonon coupling, RIXS is instead sensitive to \emph{exciton}-phonon coupling. In this case, the experimentally obtained $M_0$ actually refers to the exciton-phonon coupling self-energy.
Further theoretical work is required to conclusively determine whether this is indeed the case for correlated materials.


\begin{table}
\begin{tabular}{p{2.7cm}ccccc}
Compound & $\omega_0$ & $g$ & $\Gamma$ & $M_0$ & Absorption  \\
& (meV) &  & (eV) & (meV) &  edge\\
\hline
\hline
BaTiO$_3$ \cite{fatale2016} & 65 & 18(4) & 0.37 & 275(25) & Ti L$_3$\\
NdBa$_2$Cu$_3$O$_7$ \cite{rossi2019, braicovich2019} & 70 & 7(2) & 0.28 & 180(30) & Cu L$_3$ \\
Ca$_{2-x}$Y$_{x}$Cu$_{5}$O$_{10}$ \cite{lee2013,lee2014} & 70 & 10 & 0.15 & 220 & O K\\
Li$_2$CuO$_2$ \cite{johnston2016} & 74 & 11 & 0.15 & 240 & O K\\
3SrIrO$_3$/1SrTiO$_3$ superlattice \cite{meyers2018} & 102(3) & 7(1) & 0.22 & 260(30) & O K\\
\quad -- adjusted $\Gamma$ & 102(3) & 3.3(5) & 0.15 & 180(25) & \\
\textbf{\liiro} & & & & \\
\quad -- pure O mode & 71(3) & 14(3) & 0.24 & 260(30) & O K \\
\quad -- Ir-O mode & 18(2) & 87(6) & 0.24 & 170(20) & O K \\
\end{tabular}
\caption{Comparison of magnitudes of the EPC determined by RIXS, on a number of different samples and absorption edges. The line ``adjusted $\Gamma$'' shows how $M_0$ is sensitive to the exact value of $\Gamma$ used in the Ament model.}
\label{EPC_summary}
\end{table}
\section{Acknowledgements}
\begin{acknowledgements}
We thank Diamond Light Source for the provision of beamtime under proposal SP20569, and are grateful for experimental assistance from Thomas Rice and Peter Chang.
Work in London was supported by EPSRC (UK) (Grant No. EP/N027671, and No. EP/N034872/1). E.P. and T.S. acknowledge funding by the Swiss National Science Foundation (SNSF) through the ``Sinergia'' network Mott Physics Beyond the Heisenberg Model (MPBH) (SNSF Research Grants CRSII2\_160765/1 and CRSII2\_141962).
Work at Oxford was supported by the European Research Council (ERC) under the European Union’s Horizon 2020 research and innovation programme Grant Agreement Number 788814 (EQFT). Work at Augsburg wassupported by the German Research Foundation (DFG) via Project No. 107745057 (TRR80). 
\end{acknowledgements}

\section{Appendix}
\appendix
\section{Theory of EPC}
\label{EPC_info}
The RIXS cross-section is given formally by the Kramers-Heisenberg (KH) equation (notation from Ref.~\onlinecite{geondzhian2018}):
\begin{align}
\sigma(\omega, \omega_{\text{loss}}) &= \sum_{F}\left\vert\sum_{M}\frac{\braket{\Psi_F|\Delta_2^+|\Psi_M}\braket{\Psi_M|\Delta_1|\Psi_M}}{\omega_i - (E_M-E_I)+i\Gamma_M}\right\vert^2 \nonumber \\
&\times \delta(\omega_{\text{loss}}-(E_F-E_I)),
\end{align}
which includes a summation over many-body intermediate (M) and final (F) states, assuming that the system begins in a particular initial state (I). Here $\Gamma_M$ is the inverse lifetime of the intermediate state, $\omega_i$ is the incident photon energy, $\omega_{\text{loss}}$ the energy transfer, and $\Delta
_j$ the photon operator for the incident ($j=1$) or scattered ($j=2$) photon.

In order to determine the phonon contribution to the RIXS spectrum, one can apply the KH expression to an effective Hamiltonian with a first-order electron-phonon interaction. To simplify matters, we consider a single electronic state with energy $\epsilon_i$ interacting with a single Einstein phonon mode of energy $\omega_0$. The Hamiltonian in this case reduces to:
\begin{equation}
\mathcal{H} = \epsilon_i c_i^+c_i + \omega_0b_i^+b_i + Mc_i^+c_i(b_i^+ + b_i),
\end{equation}
where $M$ is a EPC constant, and $b_i^{(+)}$, $c_i^{(+)}$ are boson annihilation (creation) operators.

Through making a number of standard assumptions, it is possible to diagonalize this Hamiltonian by a canonical transformation. These are: that there is no electronic interaction between orbitals, that the vibrational mode does not scatter an electron between orbitals or sites, ideal harmonic ground- and excited-state potential energy surfaces, and neglecting the effect of the core-hole.
In the low-temperature limit, the phonon contribution to the RIXS cross-section becomes:
\begin{align}
\sigma(\omega_i, \omega_{\text{loss}}) &= \sum_{n_f}\left\vert \sum_{n_m} \frac{B_{n'',n'}(g)B_{n_m,0}(g)}{\omega_i - (g-n_m)\omega_0 + i\gamma_m}\right\vert^2 \nonumber \\
&\times \delta(\omega_{\text{loss}}-n_f\omega_0),
\end{align}
where $n'=\min{(n_m, n_f)}$, $n'' = \max{(n_m, n_f)}$, $g = (M/\omega_0)^2$ is the dimensionless coupling strength, and $B_{m,n}(g)$ are Franck-Condon factors given by:
\begin{equation}
B_{m,n}(g) = (-1)^m\sqrt{e^{-g}m!n!}\sum_{l=0}^n \frac{(-g)^l\sqrt{g}^{m-n}}{(n-l)!l!(m-n+l)!}.
\end{equation}
The EPC constant is related to the force constant by the relations $M = \sqrt{\frac{\hbar}{2\mu\omega}}|F|$ or $g=\frac{F^2}{2\hbar\mu\omega^3}$, where $\mu$ is the reduced mass of the oxygen atom, and $\omega$ corresponds to the excited-state vibrational frequency. 

\begin{table}
\begin{tabular}{c|ccccc}
$\theta\,(^{\circ})$ & $Q_{||}\,(\text{\AA}^{-1})$ & $h$ (r.l.u.) & $k$ (r.l.u.) & $l$ (r.l.u.) \\
\hline
15 & 0.472 & 0.386 & 0 & 0.039\\
20 & 0.451 & 0.369 & 0 & 0.075 \\
25 & 0.427 & 0.349 & 0 & 0.111 \\
30 & 0.399 & 0.326 & 0 & 0.146 \\
35 & 0.368 & 0.301 & 0 & 0.180 \\
40 & 0.335 & 0.274 & 0 & 0.213 \\
45 & 0.299 & 0.244 & 0 & 0.244 \\
50 & 0.261 & 0.213 & 0 & 0.273\\
55 & 0.220 & 0.180 & 0 & 0.300\\
60 & 0.178 & 0.146 & 0 & 0.325\\
70 & 0.091 & 0.074 & 0 & 0.367\\
80 & 0 & 0.000 & 0 & 0.398\\
90 & $-$0.091 & $-$0.074 & 0 & 0.416\\
100 & $-$0.178 & $-$0.146 & 0 & 0.423\\
110 & $-$0.261 & $-$0.213 & 0 & 0.416\\
120 & $-$0.335 & $-$0.274 & 0 & 0.396\\
130 & $-$0.399 & $-$0.326 & 0 & 0.365\\
\end{tabular}
\caption{In-plane momentum transfer $Q_{||}$ and Miller indices ($h,k,l$) calculated for the approximate experimental geometry ($E_{in} = \text{529.5~eV}$) within the DiffCalc package provided at Diamond Light Source.}
\label{hkl_table}
\end{table}

\section{X-ray absorption}\label{XAS_appendix}

\begin{figure}
\includegraphics{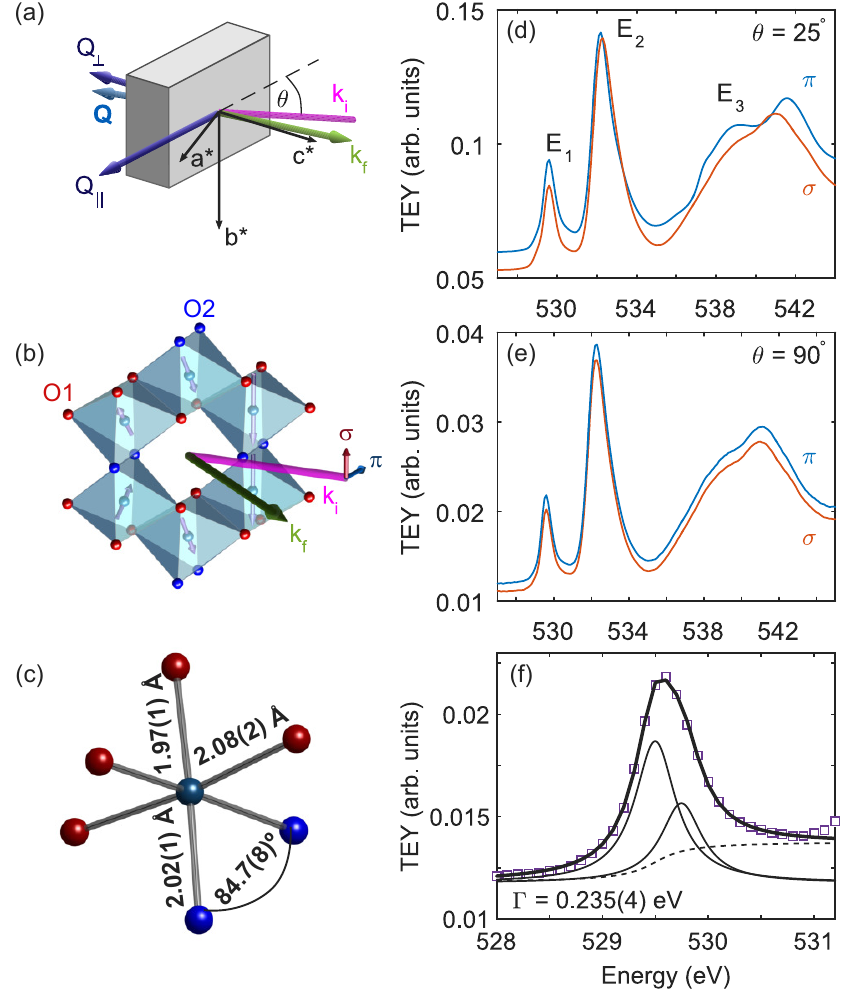}
\caption{(a): Schematic of experimental geometry, including components of momentum transfer $\bm{Q}$ that are parallel ($Q_{||}$) or perpendicular ($Q_{\perp}$) to the sample surface. (b): Orientation of incident photon polarization vectors ($\sigma$, $\pi$) with respect to the crystal structure. The two inequivalent oxygen sites (O1, O2) are highlighted.
(c): Local coordination environment for a single IrO$_6$ octahedron, annotated with bond lengths and angles. 
(d,e): XAS spectra collected on \liiro\ in total electron yield (TEY) mode for different incident polarizations at $\theta=25^{\circ}$ (a) and $\theta=90^{\circ}$ (b). (f): Fit to pre-peak at $E_1$ for $\theta=90^{\circ}$, in order to extract the inverse core-hole lifetime $\Gamma$.}
\label{XAS}
\end{figure}

X-ray absorption (XAS) measurements were performed at 10~K for incident linear horizontal (LH, $\pi$), and linear vertical (LV, $\sigma$) polarizations. Both notations shall be used interchangeably where necessary for clarity. The data are plotted in Fig.~\ref{XAS}(d--f). 
The spectra show three prominent features peaking about 529, 532, and 538~eV, labelled $E_1$, $E_2$, and $E_3$ in Figure \ref{XAS}(d) respectively.
Feature $E_3$ corresponds to transitions between Ir 6s (and higher energy orbitals) to O 2p states. The pre-edge features, $E_1$ ($E_2$), meanwhile, arise due to hybridization between the Ir $5d$ $t_{2g}$ ($e_g$) and O $2p$ states respectively.  
No significant differences in the spectra can be observed either as a function of momentum transfer, or by varying the incident polarization. 

The inverse core-hole lifetime $\Gamma$ was estimated from the XAS spectrum collected at normal incidence and $\pi$-incident polarization. This was achieved by fitting the pre-peak at $E_1$ to an arctangent function plus a Lorentzian. However, a single Lorentzian did not give a satisfactory fit to the data; two Lorentzians of equal width separated by 0.25~eV were required.
From the fit, we find that $\Gamma=\text{0.235(4)~eV}$, which is in good agreement with previous experimental estimates at the oxygen K-edge \cite{meyers2018}. We note, however, that it is significantly larger than theoretical values \cite{neeb1994, menzel1996, coreno1999, sakho2015}. It was also assumed throughout the manuscript that $\Gamma$ was constant as a function of incident energy and energy transfer \cite{ament_phd}.

In Sr$_2$IrO$_4$ and Sr$_3$Ir$_2$O$_7$, the observed splitting of peak $E_1$ was attributed to selective coupling to the in-plane or apical oxygen atoms \cite{lu2018}. This could be clearly observed from the polarization dependence of the XAS spectra. For \liiro\, there are two inequivalent oxygen sites \cite{omalley2008}, along with significant distortion of the IrO$_6$ octahedra [Fig.~\ref{XAS}(c)]. We suggest that the lack of polarization dependence for \liiro\ may be due to the experimental geometry: the IrO$_6$ octahedra are tilted by around $45^{\circ}$ out of the scattering plane [Fig.~\ref{XAS}(b)]. Meanwhile the lack of dependence upon momentum transfer may result from significant mixing between the $j=1/2$ ground state and the $j=3/2$ band, as proposed for Na$_2$IrO$_3$ \cite{sohn2013}.

\section{Self-absorption corrections}
\label{selfabs}
When performing a RIXS experiment, the incident photon energy is typically tuned to the maximum of a particular atomic resonance in order to maximise the signal. In our study, however, the incident energy was varied systematically in order to examine the effect upon the electron-phonon coupling. 
In common with References \onlinecite{minola2015} and \onlinecite{fumagalli2019}, we define a correction factor $C(\omega_1, \omega_2)$ which depends upon the incident (scattered) photon energy, such that the corrected RIXS intensity is given by:
\begin{equation}
I_{\text{corr}}(\omega_2) = \frac{I_{\text{meas}}(\omega_2)}{C(\omega_1, \omega_2)}.
\end{equation}
Specifically, we define $C$ by:
\begin{equation}
C = \frac{1}{1 + t(\omega_1, \omega_2)u}
\end{equation}
where $u=\cos{(\theta_{in})}/\!\cos{(\theta_{out})}$ is a geometrical factor depending on the angle of the incident (outgoing) beam with respect to the sample surface normal, and:
\begin{equation}
t(\omega_1, \omega_2) = \frac{\alpha_0 + \alpha_i(\omega_2)}{\alpha_0 + \alpha_i(\omega_1)},
\end{equation} 
where $\alpha_0$ and $\alpha_i$ are the non-resonant and resonant parts of the absorption coefficient. All experimental data presented in this manuscript has been corrected in this way.

\section{Effect of detuning incident photon energy}
\begin{figure}
\includegraphics{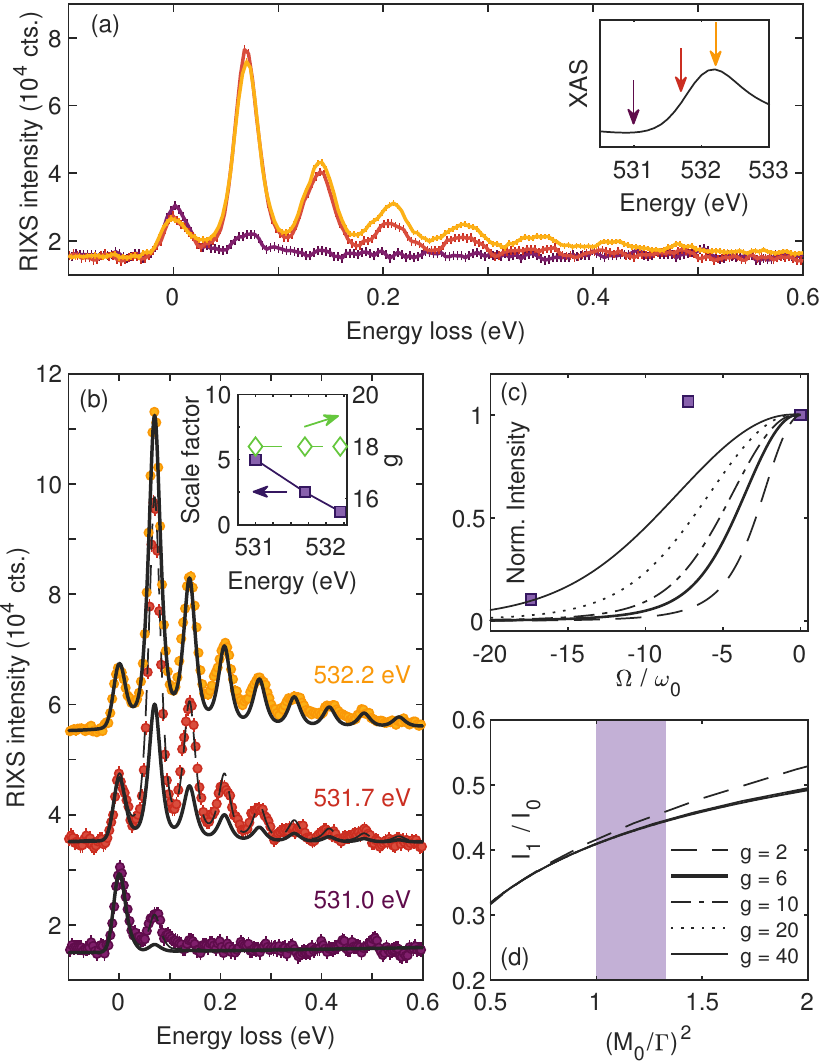}
\caption{Effect of detuning the incident photon energy away from resonance. (a): RIXS spectra collected at $\theta = 25^{\circ}$ ($Q_{||}=\text{0.42\,\AA}^{-1}$). Inset shows the incident photon energies used compared with the x-ray absorption spectrum. (b): Same spectra as plotted in (a), but overlaid with fits to the Ament model with $\omega_0 = \text{70~meV}$, $g = 18$, $\Gamma = \text{0.235~eV}$, and a scale factor which were fixed for the three different incident energies. Dashed lines are the results of fits where the relative scale factor and $g$ were allowed to vary; the fitted values are plotted in the inset.
(c): One-phonon intensity $I_0$ (squares) plotted against the dimensionless variable $\Omega / \omega_0$, where $\Omega = (E_{in} - \text{532.2~eV}$) refers to the detuning. Added are the theoretical predictions of the Ament model for different values of $g$. All data are normalized to the value at resonance.
(d): Ratio between the two-phonon and one-phonon intensities as a function of $(M_0/\Gamma)^2$. The shaded area highlights the experimentally determined region for \liiro.}
\label{EPC_detuning}
\end{figure}


In Fig.~\ref{EPC_detuning}, we plot the effect of detuning the incident photon energy away from the maximum of the resonance at 532.2~eV. This provides an alternative method of estimating the electron-phonon coupling, as detailed in Refs.~\onlinecite{rossi2019} and \onlinecite{braicovich2019}. At the Cu $L_3$ edge, this is the only viable method for estimating the EPC as any higher phonon harmonics are either very weak, or buried under spin/charge fluctuations.

In our data, we find that detuning by $-\text{1.2~eV}$ drastically reduces the intensity of the phonon harmonics, such that only a weak single-phonon peak is visible [Fig.~\ref{EPC_detuning}(a)]. Detuning by $-\text{0.5~eV}$, however, only appears to significantly affect the harmonics with $\nu>2$. Noticeably, the intensities of the $\nu=0,1$ modes appear almost unchanged with respect to those at resonance.
It was verified that this was not an effect of implementing the self-absorption correction.
The model that best described the data at resonance proved a poor fit when extending it to the energy detuned spectra [solid lines in Fig.~\ref{EPC_detuning}(b). Not only were the overall intensities of the harmonics grossly underestimated, but their relative intensities were also poorly fitted. We were only able to fully describe the data by allowing both the overall scale factor, and dimensionless EPC coupling parameter $g$, to vary as a function of $\Omega$.
In fact, it can be shown that no feasible value of $g$ can successfully model the experimentally observed intensities without varying the overall scale factor [Fig.~\ref{EPC_detuning}(c)] \footnote{As an aside, we \emph{were} able to model all the energy detuned data with a single value of $g$, but only by allowing the overall scale factor to vary, and by setting $\Omega$ to have the opposite sign.}.

There are a number of reasons why the energy detuning analysis does not work as well at the oxygen $K$-edge compared to the copper $L_3$ edge.
In practice, the assumption that the effect of the core-hole can be neglected starts to become invalid for systems with sufficiently large EPC, and small inverse core-hole lifetime $\Gamma$.
This can be seen in Fig.~\ref{EPC_detuning}(d), where the ratio of the two-phonon to one-phonon intensity is plotted for different $g$ as a function of the dimensionless variable $(M_0/\Gamma)^2$. For small $(M_0/\Gamma)^2$, all the curves appear to superimpose on top of each other, giving rise to a well-defined single-valued problem.
Yet these curves start to diverge for $(M_0/\Gamma)^2>1$, especially for small values of $g$. This means that there may not be a unique solution.
In the cuprates measured thus far, the condition that $(M_0/\Gamma)^2 < 1$ always appears to be satisfied since $\Gamma = \text{0.28~eV}$ at the Cu $L_3$ edge. For \liiro, however, we find that $1 \leq (M_0/\Gamma)^2 \leq 1.35$ for the phonon modes measured. 

Another complication (as discussed previously) is that each multi-phonon peak may in fact be comprised of more than one overlapping excitation. It is entirely possible that these modes have different EPC strengths, and even different resonant behavior. If this is the case, then the analysis presented in this manuscript would represent an average coupling strength, and may explain some of the discrepancy from theory upon detuning.

\section{Momentum dependence of orbital excitations}
\begin{figure}
\includegraphics{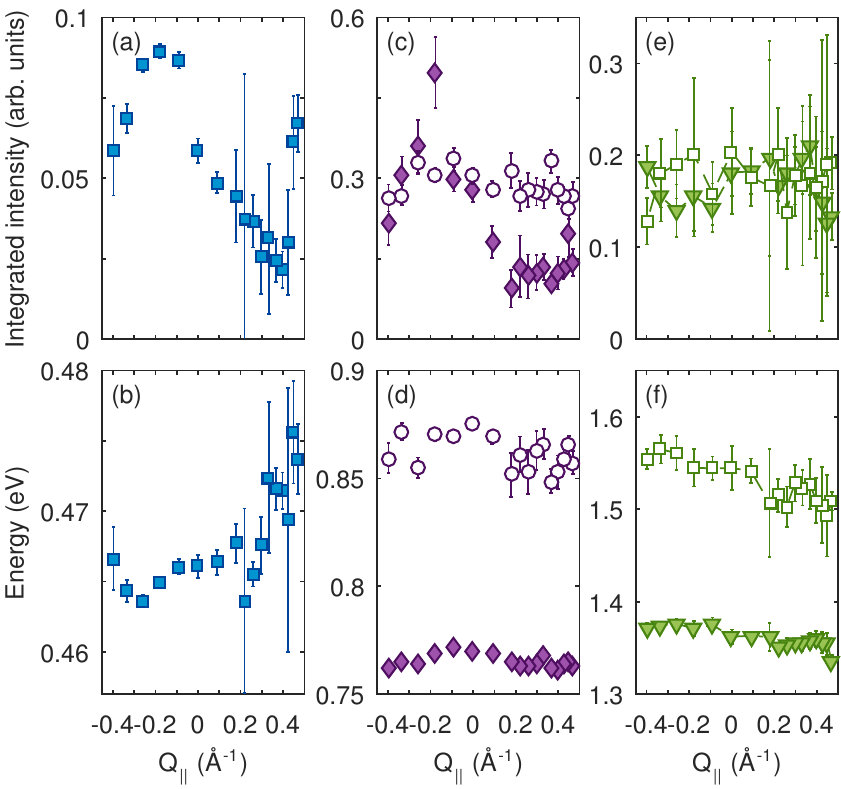}
\caption{Fitted integrated intensity and energy of: spin-orbit exciton (a,b), local $j_{3/2}\leftarrow j_{1/2}$ (c,d), and intersite (e,f) excitations. The colors of the data points are the same as the excitations shaded in Fig.~5(a) of the main text; panel (d) is a reproduction of Fig.~5(d) in the main text.}
\label{orbital_Qdep_full}
\end{figure}

In Fig.~\ref{orbital_Qdep_full}, we plot the intensities and energies of the various orbital excitations as a function of in-plane momentum transfer $Q_{||}$. The intensity variation as a function of $Q_{||}$ is predominantly due to polarization effects which enter into the RIXS cross-section. This effect has been studied in some detail by Kang et al \cite{kang2019}, and shall not be discussed further here.

The excitation energies, meanwhile, appear to be weakly dispersive. This was discussed for the local $j_{3/2}\leftarrow j_{1/2}$ transitions in the main text. The \emph{intersite} $j_{3/2}\leftarrow j_{1/2}$ excitations [Fig.~\ref{orbital_Qdep_full}(f)] seem to disperse even more strongly than those confined to a single site \cite{hermann2017}. Note, however, that these excitations are rather broad and overlap significantly. This also explains the large errorbars in the intensity as shown in Fig.~\ref{orbital_Qdep_full}(e).
We comment that these intersite transitions are absent in the Ir L$_3$ RIXS data \cite{gretarsson2013}, but have been previously observed by optical conductivity measurements \cite{hermann2017}.
Since these processes involve hopping via an intermediate oxygen atom, one may expect a greater sensitivity to them at the oxygen K-edge. Indeed, such behavior has already been demonstrated for YVO$_3$ \cite{benckiser2013}.

Finally we discuss the spin-orbit exciton. In Sr$_2$IrO$_4$, the dispersion of this feature manifests directly from the fact that the propagating exciton creates a string of flipped isospins along its hopping path \cite{kim2012_sr214, kim2014, clancy2019}. The corresponding bandwidth is on the order of $2J$, and can therefore provide an indirect method of determining the magnitude of exchange coupling.
In \liiro, we observe that the spin-orbit exciton has a bandwidth of ca.~10~meV within the experimentally accessible region of $Q_{||}$. This implies an effective exchange parameter $J\sim \text{5~meV}$, which is broadly consistent with the low energy spin excitations observed in this material by inelastic neutron scattering \cite{choi2019}.

%

\end{document}